\documentclass[sigconf]{acmart}
\AtBeginDocument{%
  }

\setcopyright{acmlicensed}
\copyrightyear{2018}
\acmYear{2018}
\acmDOI{XXXXXXX.XXXXXXX}
\acmConference[Conference acronym 'XX]{Make sure to enter the correct
  conference title from your rights confirmation email}{June 03--05,
  2018}{Woodstock, NY}
\acmISBN{978-1-4503-XXXX-X/2018/06}

\begin{document}

\title{A Technical Report on the Second Place Solution for the CIKM 2025 AnalytiCup Competition}

\author{Haotao Xie}
\orcid{0009-0002-3061-8090}
\affiliation{%
  \institution{The Hangzhou International Innovation Institute}
  \institution{Beihang University}
  \city{HangZhou}
  \state{Zhejiang}
  \country{China}
}
\email{haotaoxie@buaa.edu.cn}

\author{Ruilin Chen}
\affiliation{%
  \institution{School of Mechanical and 
Automotive Engineering}
  \institution{South China University of Technology}
  \city{GuangZhou}
  \state{Guangdong}
  \country{China}}
\email{15323544020@163.com}

\author{Yicheng Wu}
\affiliation{%
  \institution{The Hangzhou International Innovation Institute}
  \institution{Beihang University}
  \city{HangZhou}
  \state{Zhejiang}
  \country{China}}
\email{wuyicheng@buaa.edu.cn}

\author{Zhan Zhao}
\affiliation{%
  \institution{The Hangzhou International Innovation Institute}
  \institution{Beihang University}
  \city{HangZhou}
  \state{Zhejiang}
  \country{China}}
\email{zhaozhan@buaa.edu.cn}

\author{Yuanyuan Liu}
\affiliation{%
  \institution{The Hangzhou International Innovation Institute}
  \institution{Beihang University}
  \city{HangZhou}
  \state{Zhejiang}
  \country{China}}
\email{yy_liu@buaa.edu.cn}

\renewcommand{\shortauthors}{Trovato et al.}

\begin{abstract}
In this work, we address the challenge of multilingual category relevance judgment in e-commerce search, where traditional ensemble-based systems improve accuracy but at the cost of heavy training, inference, and maintenance complexity. 
To overcome this limitation, we propose a simplified yet effective framework that leverages Prompt engineering with Chain-of-Thought task decomposition to guide reasoning within a single large language model. Specifically, our approach decomposes the relevance judgment process into four interpretable subtasks: translation, intent understanding, category matching, and relevance judgment—and fine-tunes a base model (\textbf{Qwen2.5-14B}) using Low-Rank Adaptation (\textbf{LoRA}) for efficient adaptation. This design not only reduces computational and storage overhead but also enhances interpretability by explicitly structuring the model’s reasoning path. Experimental results show that our single-model framework achieves competitive accuracy and high inference efficiency, processing \textbf{20} samples per second on a single A100 GPU.
In the CIKM 2025 AnalytiCup Competition Proposals, our method achieved \textbf{0.8902} on the public leaderboard and \textbf{0.8889} on the private leaderboard, validating the effectiveness and robustness of the proposed approach. These results highlight that structured prompting combined with lightweight fine-tuning can outperform complex ensemble systems, offering a new paradigm for scalable industrial AI applications.
\end{abstract}



\begin{teaserfigure}
\centering
  \includegraphics[trim=0 350 80 60, 
    clip, 
    width=\textwidth]{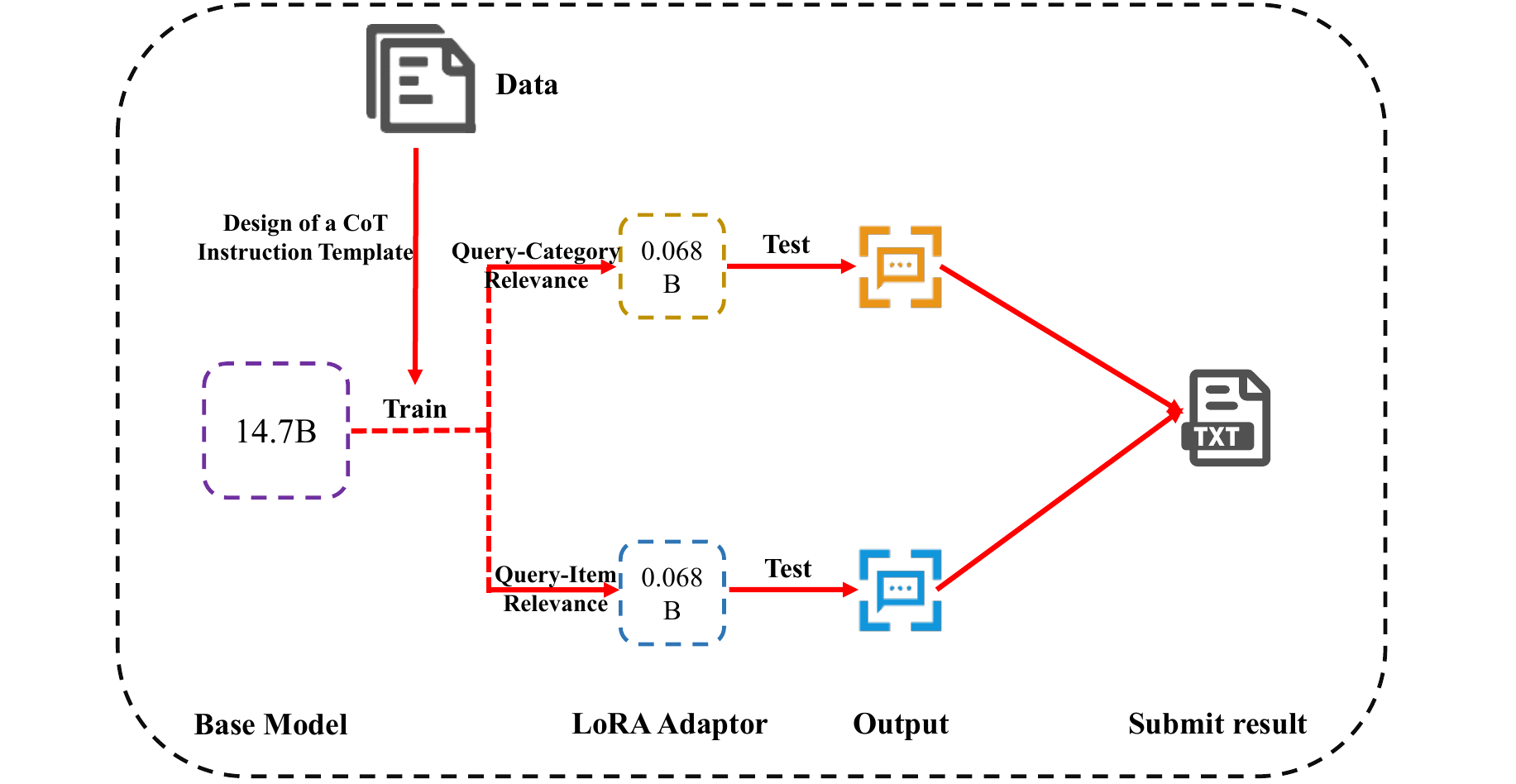}
  \caption{Overall framework of the proposed Chain-of-Thought and LoRA-based system for multilingual category relevance judgment.}
  \Description{Enjoying the baseball game from the third-base
  seats. Ichiro Suzuki preparing to bat.}
  \label{fig:teaser}
\end{teaserfigure}

\maketitle

\section{Introduction}

In the rapidly expanding field of e-commerce, accurately determining the semantic relevance between a user query and a product category is essential for improving search quality, recommendation precision, and user experience. 
This task underpins a range of downstream applications, including category classification, product retrieval, and relevance ranking, forming a core component of large-scale industrial search systems. 
However, this problem becomes particularly challenging in multilingual and cross-domain environments, where the same intent can be expressed through diverse linguistic forms, regional conventions, and hierarchical category structures. 
Subtle variations in translation, attribute description, and product taxonomy can lead to mismatches that traditional text-matching or embedding-based methods fail to capture effectively. 

Existing approaches often resort to ensemble architectures or rule-based post-processing to achieve higher accuracy. 
While these methods can reduce model variance, they introduce significant computational overhead and maintenance burden, especially when deployed across multiple languages and markets. 
More importantly, they lack interpretability—ensemble scores do not reveal the reasoning process behind a model’s decision, making it difficult to diagnose errors or generalize to unseen contexts. 
Meanwhile, although large language models (LLMs) have demonstrated strong generalization in text understanding, their reasoning process remains largely implicit, leading to unstable or inconsistent outputs across languages and domains.

To address these challenges, we propose a simple yet powerful framework that leverages \textbf{Prompt Engineering with Chain-of-Thought (CoT)\cite{wei2023chainofthoughtpromptingelicitsreasoning} task decomposition} to guide LLMs toward explicit and interpretable reasoning in the multilingual category relevance judgment task. 
Our key insight is that a well-designed prompt structure can serve as an implicit reasoning scaffold, allowing a single model to emulate the interpretability and robustness typically achieved by multi-model ensembles. 
Instead of relying on multiple specialized systems, we decompose the task into four coherent subtasks—\textit{translation}, \textit{intent understanding}, \textit{category matching}, and \textit{relevance judgment}. 
This decomposition aligns the model’s reasoning process with human decision logic, enabling it to progressively refine its understanding of semantic consistency between queries and category hierarchies. 
Such a structured reasoning paradigm transforms a black-box decision into a transparent multi-step inference process.

To efficiently adapt the model to the competition setting, we integrate \textbf{Low-Rank Adaptation (LoRA)}~\cite{DBLP:journals/corr/abs-2106-09685} for parameter-efficient fine-tuning. 
LoRA introduces trainable low-rank matrices into selected layers of the transformer, significantly reducing training cost and memory footprint while preserving generalization ability. 
This enables us to fine-tune a large base model (Qwen2.5-14B) under constrained hardware resources—achieving high accuracy and fast inference without compromising interpretability. 
Together, the CoT-based structured prompting and LoRA-based fine-tuning form a cohesive pipeline that balances simplicity, performance, and scalability.

Extensive experiments conducted on the \textit{CIKM 2025 AnalytiCup Competition Proposals} verify the effectiveness and robustness of our method. 
Our single-model system achieves a score of \textbf{0.8902} on the public leaderboard and \textbf{0.8889} on the private leaderboard, outperforming traditional ensemble-based baselines. 
In addition to strong empirical results, our framework demonstrates high inference throughput—processing 20 samples per second on a single NVIDIA A100 GPU—making it well-suited for real-world deployment in large-scale e-commerce systems. 
These results confirm that explicit reasoning, when coupled with lightweight adaptation, can substitute for architectural complexity without sacrificing accuracy, efficiency, or scalability.

In summary, our main contributions are as follows:
\begin{itemize}
    \item We analyze the limitations of ensemble-based multilingual relevance judgment systems and reformulate the problem as a structured reasoning task within a single large language model.
    \item We propose a unified framework combining \textbf{CoT-guided prompt engineering} and \textbf{LoRA-based fine-tuning}, achieving a balance between interpretability, computational efficiency, and accuracy.
    \item We conduct extensive experiments on a large-scale benchmark, providing empirical evidence that structured prompting can serve as a general principle for efficient and deployable industrial AI systems.
    \item We demonstrate the practical feasibility of our approach through successful participation in the \textit{CIKM 2025 AnalytiCup Competition Proposals}, achieving state-of-the-art results under constrained resources.
\end{itemize}

\section{Preliminaries}

\subsection{Prompt Engineering}

Prompt engineering serves as the foundation of our framework, providing a structured interface between large language models (LLMs) and downstream e-commerce relevance judgment tasks. In this work, we adopt a task decomposition strategy that transforms a complex, multilingual classification problem into a sequence of interpretable sub-tasks. Specifically, the overall task of category relevance judgment is divided into four steps: \textit{translation}, \textit{intent understanding}, \textit{category matching}, and \textit{relevance judgment}. 

Such decomposition follows the principle of Chain-of-Thought (CoT) prompting, which encourages the model to reason explicitly through intermediate steps rather than relying on latent representations. For each sub-task, we design targeted instruction templates that align with the model’s internal reasoning process, ensuring that the generated intermediate outputs can be seamlessly integrated into subsequent stages. This structured prompt design not only enhances interpretability and controllability but also stabilizes model behavior across different languages and product categories.

Compared to traditional flat prompting, our hierarchical design enables the model to progressively refine its understanding of user intent and product semantics, thus mitigating common errors caused by cross-lingual ambiguity and inconsistent category hierarchies. The resulting prompts form the foundation of our fine-tuning dataset, which is subsequently used to adapt the base model via parameter-efficient learning techniques such as LoRA.

\subsection{Low-Rank Adaptation}

To efficiently adapt large-scale language models to task-specific objectives, we employ \textit{Low-Rank Adaptation (LoRA)}, a parameter-efficient fine-tuning technique that introduces trainable low-rank matrices into selected layers of the transformer architecture. Instead of updating all model parameters, LoRA learns a small number of additional parameters that approximate the necessary weight adjustments in a low-dimensional subspace. This significantly reduces memory usage and computational cost while preserving the pre-trained model’s generalization capability.

Formally, for a given weight matrix $W \in \mathbb{R}^{d \times k}$ in the transformer layers, LoRA re-parameterizes it as:
\[
W' = W + \Delta W = W + BA,
\]
where $A \in \mathbb{R}^{r \times k}$ and $B \in \mathbb{R}^{d \times r}$ are low-rank matrices with $r \ll \min(d, k)$. Only $A$ and $B$ are updated during fine-tuning, and the original weights $W$ remain frozen. 

In our implementation, LoRA modules are applied to the \textit{q\_proj}, \textit{k\_proj}, \textit{v\_proj}, \textit{o\_proj}, \textit{gate\_proj}, \textit{up\_proj}, and \textit{down\_proj} layers to enhance both the attention mechanism and the feed-forward network’s expressiveness. We set the rank $r=24$, the scaling factor $\alpha=32$, and employ a dropout rate of 0.1 to prevent overfitting. 

This design achieves an optimal balance between efficiency and effectiveness: the model can be trained on a single NVIDIA A100 (40GB) GPU while retaining high performance across multilingual test sets. Moreover, by combining LoRA with structured prompting, we ensure that the adaptation process preserves the reasoning structure learned from the prompt decomposition, thereby maintaining interpretability and stability during inference.

\begin{figure}[h]
  \centering
  \includegraphics[trim=128 130 161 103, 
    clip, 
    width=\linewidth]{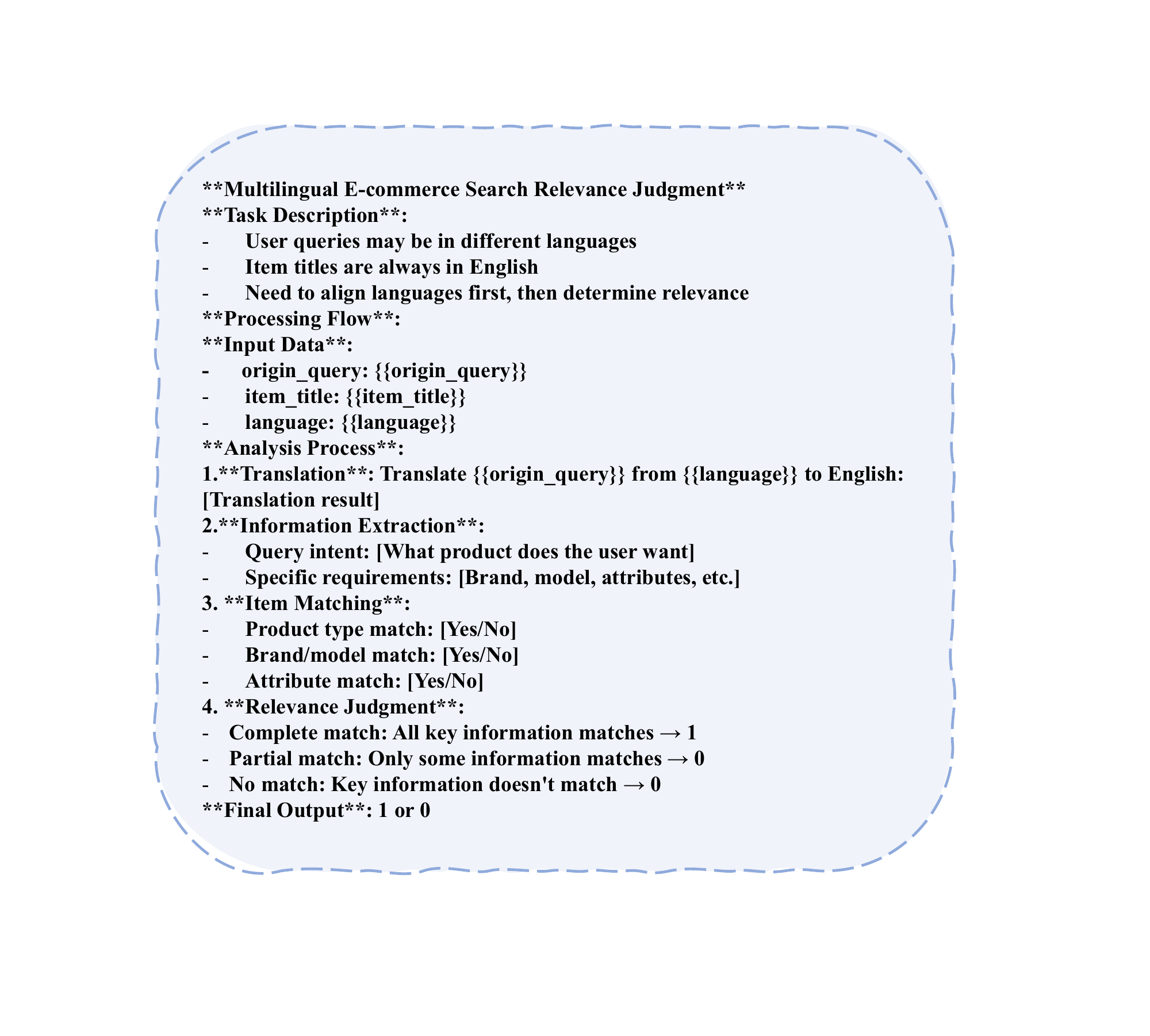}
  \caption{Prompt template design for the Query-Item Relevance (QI) task.}
  \Description{}
\end{figure}

\section{Method}

\subsection{Overall Framework}

Our goal is to build an efficient and interpretable system for multilingual category relevance judgment in e-commerce search. 
Unlike previous approaches that rely on complex ensemble systems, our framework is designed to achieve competitive performance using a single large language model (LLM). 
The overall architecture, illustrated in Figure~\ref{fig:teaser}, consists of three key stages: (1) task decomposition and prompt construction, (2) parameter-efficient fine-tuning via LoRA, and (3) high-throughput inference.

Given a multilingual query–category pair, the model first performs a stepwise reasoning process guided by structured prompts. Each intermediate result—such as translation, intent understanding, or category comparison—provides explicit reasoning signals that enhance final decision quality. The entire process follows an ``explicit reasoning within a single model'' paradigm, which replaces the need for multiple specialized models typically used in ensemble systems.

\subsection{Task Decomposition and Prompt Construction}

To enable structured reasoning, we reformulate the category relevance judgment task as a four-step pipeline following the principle of Chain-of-Thought (CoT) prompting:
\begin{enumerate}
    \item \textbf{Translation:} Convert multilingual user queries into English to reduce cross-lingual interference and unify semantic space.
    \item \textbf{Intent Understanding:} Extract user intent, product type, and key attributes from the translated query.
    \item \textbf{Category Matching:} Compare the extracted semantic representation with candidate category paths from three perspectives—type consistency, hierarchical alignment, and attribute compatibility.
    \item \textbf{Relevance Judgment:} Integrate the previous reasoning results and output a binary label indicating relevance (1) or irrelevance (0).
\end{enumerate}

For each step, we design task-specific instruction templates to guide the model’s reasoning process. 
For instance, the translation stage adopts a direct instruction format (“Translate the following query into English: ...”), while the intent understanding and category matching stages use question–answer style prompts to encourage structured outputs.
All subtasks are merged into a unified instruction–response dataset, forming the foundation for LoRA-based fine-tuning.

\subsection{LoRA-based Fine-tuning}

We employ LoRA to efficiently fine-tune the base model, Qwen2.5-14B. 
By introducing trainable low-rank matrices into the attention and feed-forward layers, LoRA drastically reduces the number of trainable parameters while maintaining expressiveness.
In practice, LoRA adapters are injected into the \textit{q\_proj}, \textit{k\_proj}, \textit{v\_proj}, \textit{o\_proj}, \textit{gate\_proj}, \textit{up\_proj}, and \textit{down\_proj} modules, with rank $r=24$, scaling factor $\alpha=32$, and dropout rate of 0.1.
The model is trained on a single NVIDIA A100 (40GB) GPU with a batch size of 8, gradient accumulation of 2, and learning rate of 2e-4.
This setup enables us to achieve high-quality fine-tuning under constrained computational resources.

\subsection{Inference and Efficiency}

During inference, the fine-tuned model performs reasoning in a single forward pass guided by the structured prompt.
Thanks to the CoT-based decomposition, each decision step is explicit and verifiable, allowing for transparent evaluation and debugging.
Our optimized inference pipeline achieves an average throughput of 20 samples per second on a single A100 GPU when processing 100,000 query–category pairs, 
demonstrating that structured prompting combined with LoRA fine-tuning not only preserves interpretability but also enables scalable industrial deployment.

\subsection{Summary of Advantages}

Our method offers three key advantages:
(1) \textbf{Simplicity:} a single LLM replaces the need for multi-model ensembles;
(2) \textbf{Interpretability:} the CoT-based decomposition exposes internal reasoning steps;
(3) \textbf{Efficiency:} LoRA fine-tuning and lightweight inference ensure practicality for large-scale industrial applications.
Together, these design choices enable a balance between accuracy, efficiency, and interpretability, validated by our strong performance in the \textit{CIKM 2025 AnalytiCup Competition Proposals}—scoring 0.8902 on the public leaderboard and 0.8889 on the private leaderboard.

\begin{figure}[h]
  \centering
  \includegraphics[trim=130 72 162 0, 
    clip, 
    width=\linewidth]{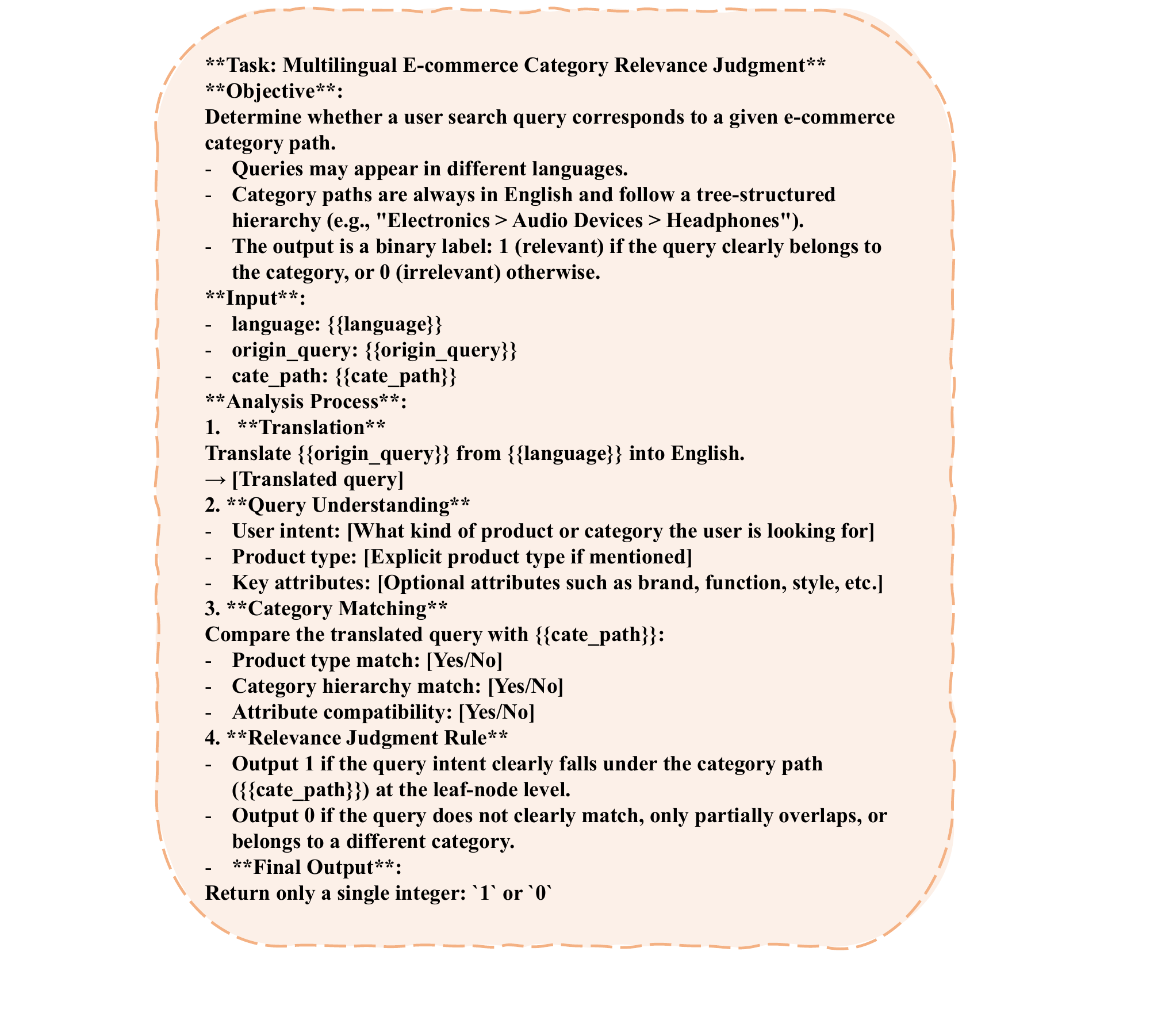}
  \caption{Prompt template design for the Query-Category Relevance (QC) task.}
  \Description{}
\end{figure}

\section{Experiments}

\subsection{Dataset and Task Setup}

The evaluation is conducted under the setting of the \textit{CIKM 2025 AnalytiCup Competition Proposals}, focusing on the multilingual category relevance judgment task. 
The dataset consists of query–category pairs from multiple languages, where each query may contain product names, attributes, or user intents expressed in different linguistic forms.
Each pair is labeled as either \textit{relevant} (1) or \textit{irrelevant} (0) based on the semantic alignment between the query and the target category path.

Before training, we perform data preprocessing that includes:  
(1) language completion by mapping language codes (e.g., “en”, “es”, “fr”, “ar”) to their corresponding language names;  
(2) translation of all non-English queries into English to normalize the input space;  
(3) prompt construction according to the four-step decomposition.  
The resulting instruction–response dataset provides structured supervision for LoRA fine-tuning.

\subsection{Experimental Configuration}

All experiments are conducted on a single NVIDIA A100-PCIE-40GB GPU, covering both training and inference. 
We use the \textbf{Qwen2.5-14B} base model instead of the instruction-tuned variant to maintain neutrality and flexibility in downstream adaptation. 
Fine-tuning is performed using the LoRA framework with rank $r=24$, scaling factor $\alpha=32$, and dropout 0.1.  
Adapters are inserted into \textit{q\_proj}, \textit{k\_proj}, \textit{v\_proj}, \textit{o\_proj}, \textit{gate\_proj}, \textit{up\_proj}, and \textit{down\_proj} layers.  
The training batch size is 8, with gradient accumulation of 2, and learning rate 2e-4 for one epoch.  
Gradient checkpointing is enabled to reduce memory consumption.

During inference, the model processes 100,000 test samples in approximately 5,000 seconds, achieving an average speed of \textbf{20 samples per second}. 
All results are reported on the official competition evaluation server to ensure comparability and reproducibility.

\subsection{Results and Performance Analysis}

Table~\ref{tab:results} summarizes our performance on both public (A track) and private (B track) leaderboards. 
Our single-model framework achieves a score of \textbf{0.8902} on the A leaderboard and \textbf{0.8889} on the B leaderboard, demonstrating competitive accuracy while maintaining a lightweight architecture.

\begin{table}[h]
\centering
\caption{Performance on the CIKM 2025 AnalytiCup leaderboards.}
\begin{tabular}{lcc}
\toprule
\textbf{Method} & \textbf{A leaderboard} & \textbf{B leaderboard} \\
\midrule
Baseline & 0.8698 & ***** \\
Ours (CoT + LoRA) & \textbf{0.8902} & \textbf{0.8889} \\
\bottomrule
\end{tabular}
\label{tab:results}
\end{table}

The results show that a single large model, when guided by structured prompting and fine-tuned via LoRA, can outperform traditional methods. 
This validates our key insight: \textit{explicit reasoning and lightweight adaptation can substitute for architectural complexity without sacrificing performance.}

\section{Conclusion and Future Work}

In this work, we present a simplified yet effective framework for multilingual category relevance judgment in e-commerce search. 
Departing from traditional ensemble-based systems, our approach demonstrates that a single large language model, when equipped with structured prompting and parameter-efficient fine-tuning, can achieve competitive or even superior performance with significantly reduced complexity. 
Through Chain-of-Thought (CoT) task decomposition, the model explicitly reasons over translation, intent understanding, category matching, and relevance judgment, resulting in improved interpretability and decision consistency.

By integrating Low-Rank Adaptation (LoRA) into the fine-tuning process, we further enhance model adaptability under limited computational resources. 
Comprehensive experiments on the \textit{CIKM 2025 AnalytiCup Competition Proposals} verify the effectiveness of our approach, where the proposed single-model framework achieves a score of 0.8902 on the public leaderboard and 0.8889 on the private leaderboard. 
These results confirm that structured prompting combined with lightweight fine-tuning offers a strong balance between performance, efficiency, and interpretability for industrial-scale AI applications.

In future work, we plan to extend our framework in three directions.  
First, we aim to explore \textbf{cross-domain generalization}, adapting the model to other e-commerce tasks such as product attribute extraction and query rewriting.  
Second, we will investigate \textbf{multimodal integration}, incorporating image and textual cues to enhance semantic understanding in product-level reasoning.  
Finally, we intend to study \textbf{reinforcement-based prompt optimization}, enabling adaptive prompt refinement guided by feedback from downstream performance metrics.  
We believe these directions will further validate the scalability of our approach and contribute to the broader development of efficient, interpretable, and deployable large-model solutions in industrial AI.

\begin{acks}
This work was carried out as part of the \textbf{CIKM 2025 AnalytiCup Competition Proposals}, organized by \textbf{Alibaba International Tech}. We would like to express our sincere gratitude to the organizers for providing the competition platform, datasets, and technical support that made this research possible.
\end{acks}

\bibliographystyle{ACM-Reference-Format}
\bibliography{sample-base}

\end{document}